# A Dynamic Algebraic Specification for Social Networks

Katerina Ksystra[1], Konstantinos Barlas[1], Nikolaos Triantafyllou[1] and Petros Stefaneas[2],

[1] School of Electrical and Computer Engineering, National Technical University of Athens,
Heroon Polytechniou 9, 15780 Zografou, Athens, Greece
{nitriant, kosbarl, katksy}@central.ntua.gr
[2] School of Applied Mathematical and Physical Sciences, National Technical University of
Athens, Heroon Polytechniou 9, 15780 Zografou, Athens, Greece
petros@math.ntua.gr

**Abstract.** With the help of the Internet, social networks have grown rapidly. This has increased security requirements. We present a formalization of social networks as composite behavioral objects, defined using the Observational Transition System (OTS) approach. Our definition is then translated to the OTS/CafeOBJ algebraic specification methodology. This translation allows the formal verification of safety properties for social networks via the Proof Score method. Finally, using this methodology we formally verify some security properties.

## 1 Introduction

A Social Network is a very broad term and can usually be defined in any of the following three ways:

*Social Network as a set of relationships*: More formally, it contains a set of objects (nodes, representing individuals) and a mapping or description of relations (usually representing types of interdependency such as friendship or common interests) between the object or nodes. [16]

*Social Networking Service as an online representation*: consists of a representation of a user (via a profile), his/her connections to other profiles, social links, and a variety of additional services.

*Social Networking Sites as web-based services*: they allow individuals to construct a public or semi-public profile within a bounded system, articulate a list of other users with whom they share a connection, and view and traverse their list of connections and those made by others within the system. The nature and nomenclature of these connections may vary from site to site [17].

While the concept of social networks originates back to the 19[th] century (a social network can represent various concepts, such as the relationship of a teacher and his/her students, a family, etc.), it's only recently received a huge popularity boost with the help of the internet. Nowadays, the strain that the millions of users put on the structures of those (online) services is sometimes difficult to handle. Due to the problems arising with this increase of use, several attempts have been made over the last years to describe and analyze the structure and properties of a social network as a formal concept. Our approach is based on formal methods, mathematically-based techniques that are used in the specification, development and verification of software and hardware systems.

Papers related to the mathematical modeling of Social Networks include [14, 15]. [14] attempts to visualize and reduce the size of social networks with the help of formal concept analysis. It regards a social network as a static structure, examining snapshots (taken at a certain moment) and produce results based on that instance, but do not consider the network evolution in their approach. In [15] the authors suggest that formal methods can provide a logical foundation in order to express and enforce privacy and security policies on social networks, however the specification done is quite minimal and its extension relies on external tools such as Java. Our approach differs in the sense that we regard the social network as a dynamic composition of behavioral objects, specifically as profiles that are connected through friendship relationships and that the network evolves by adding or deleting profiles.

The rest of the paper is organized as follows: Section 2 describes the mathematical background for observational transition systems (OTS), a quick introduction to the theory behavioral object composition and an abstract definition of a social network as an OTS. Section 3 provides an overview of the CafeOBJ specification language/system and the implementation of OTSs in this framework. Section 4 describes our modeling proposal while section 5 demonstrates how critical safety properties of our system can be verified. Finally, section 6 concludes our paper with some future goals.

## 2 A Social network as an OTS

An *Observational Transition System*, or *OTS* [12, 13], is a transition system that can be written in terms of equations. We assume that there exists a universal states space $Y$ and that each data type we need to use in our OTS, including their equivalence relationship, has been declared in advance. An OTS $S$ is defined as the triplet $<O, I, T>$ where:

1. $O$ is a finite set of observers. Each $o \in O$ is a function $o : Y \to D$, where D is a data type that may differ from observer to observer. Given an OTS $S$ and two states $u_1, u_2 \in Y$, the equivalence ($u_1 =_S u_2$) between them with regards to (wrt) $S$ is defined as $\forall o \in O, o(u_1) = o(u_2)$.
2. $I$ is the set of initial states such that $I \subseteq Y$.
3. $T$ is a set of conditional transitions. Each $\tau \in T$ is a function $\tau : Y \to Y$, such that $\tau(u_1) = \tau(u_2)$ for each $u_1, u_2 \in Y/=_S$. For each $u \in Y, \tau(u)$ is called the successor state of u wrt $\tau$. The condition $c_\tau$ of $\tau$ is called the effective condition. Also, for each $u \in Y, \tau(u) = u$ if $not\, c_\tau(u)$.

Observers and transitions may be parameterized. Generally observers and transitions are denoted as $o_{i1,...,im}$ and $\tau_{j1,...,jn}$ respectively, provided that $m, n \geq 0$ and that there exist data types $D_k, D$ where $k = i_1, ..., i_m, j_1, ..., j_n$.

As defined in [1], a *Behavioral Object* is, informally, a behavioral specification that denotes formally the state space of the object, together with a set of actions (transitions) that change the state, as well as a set of observers that return the values of the data types that interest us. It is clear that the above definition of an OTS complies with this informal definition of a behavioral object.

The semantics of behavioral specification is based on *Hidden Algebra* [4, 7, 8] which is a refinement of general many-sorted algebra [1]. In a nutshell, Hidden Algebra extends ordinary general algebra with extra sorts representing the "states" of an object or an abstract machine. Due to this addition, we can represent in a natural way the equality of two states of a machine, by introducing a new satisfaction of algebras and sentences called behavioral satisfaction [1].

*Behavioral Object Composition* methodology has been defined formally in [1] so that it can be exported to any specification and verification language that implements a behavioral logic. The behavioral object composition methodology is *hierarchical* since the composition of behavioral objects yields another behavioral object, which can be used further for another composition.

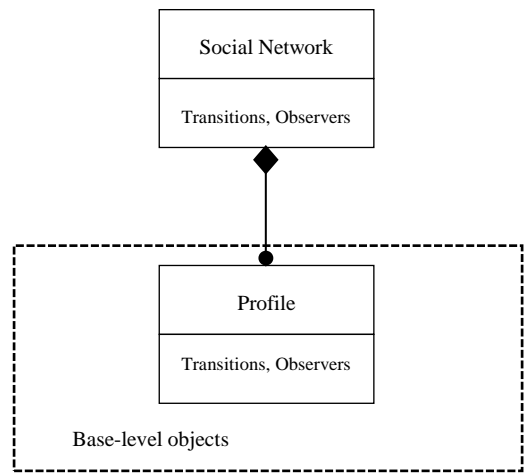

Fig. 1. Hierarchical behavioral object composition in UML notation for the Social Network OTS

The non-compound objects (i.e. objects with no components) are called *base-level objects*. A composition is represented in UML [11] by lines tipped by diamonds, and if necessary, qualified by the numbers of components (**1** for one and **\*** for many). Also, the circle on the top of base-line object denotes that the component object contains arbitrary many such components, i.e. the circle denotes a dynamic object.

The main technical concept of the behavioral object composition is the *Projection Operators* from the states (hidden sort) of the composite object to the states of the component objects. These are represented above as lines in the UML diagram and are subject to the following rules [1]:

1. All actions of the components are represented by actions at the level of the compound object via projection operations.
2. Each observation of the compound object is related via the projection operations to an observation of some component.
3. Each constant state of the compound object is projected to a constant state on each component.

The connection between the compound object and the components happens at the level of the compound object. This means that the equations relating the actions and the observations of the compound object to those of the components are declared in the specification of the compound object.

There are several ways to compose an object from component objects. In respect to synchronization we have *Parallel* composition, if the changes on the states of the object do not affect the states of the other objects in the same level and *Parallel with Synchronization* when the changes in the state of one object may alter the state of an object in the same level. Also in respect to the number of objects that compose a composite object, we have *Dynamic Composition* if that number is not fixed else the composition is *Static*. Thus, the most general case to create a composite object is *Dynamic Synchronized Parallel Composition*. For example, the specification of a bank account system that allows transfers of funds between accounts and also allows for new accounts to be created or for old to be deleted is most naturally specified as a dynamic synchronized parallel composition of the bank accounts specifications.

Our specification of Social networks follows the path of Dynamic Synchronized Parallel Composition. As *base-level object* we have OTSs $< O, I, T >$ that represent a user of a social network as a *Profile* (this notion will be better explained in section 4). The social network itself is specified as a *Composite OTS*, $< O', I', T' >$, where $I' \subseteq Y'$ and $T' = T'' \cup \{a, d\}$, with $a$ and $d$ two special transitions that either add or delete a profile from our Social Network OTS. Finally, $O' = O'' \cup \{p\}$, with $p$ being a parameterized projection operator that takes us from the state of the social network to the state of the corresponding profile. This is formally defined as $p : Y' \to Y$, with $p(X_{i1}, \ldots, X_{in}, H') = H$. In the latter equation, $X_{ik}$ denotes the required visible sorts, $H'$ denotes a state of the composite object and $H$ a state of the base level objects. The diagram of figure 1 describes all the above.

## 3   Social OTS in CafeOBJ

CafeOBJ is an algebraic specification language [3, 5]. We have chosen this language for our specification because an OTS can be written in CafeOBJ in a natural way. Moreover, the hierarchical object composition, based on behavioral specification presented in section 2, has already been defined in [3, 9, 6] with the use of CafeOBJ.

However, other algebraic specification languages could have been used as well, such as CASL [2] or Maude [10].

The universal state space $Y$ is denoted in CafeOBJ by a hidden sort, while each observer by an observation operator. Assuming visible sorts $V_{ij}$ and $V$ that correspond to the data types $D_k$ and $D$, where $k = i_1, \ldots, i_m$, the observation operator denoting $o_{i1,\ldots,im}$ is declared as follows; $bop\ o : V_{i1} \ldots V_{im}\ H \to V$. Any state in $I$ is denoted by a constant, say *init*, which is declared as; $op\ init : \to H$. A transition $\tau_{j1,\ldots,jn} \in T$ is denoted by a CafeOBJ action operator as follows; $bop\ \tau : V_{j1} \ldots V_{jm}\ H \to H$, with $V_k$ a visible sort corresponding to the data type $D_k$ and $k = j_1, \ldots, j_n$.

Each transition is defined by describing what the value, returned by each observer in the successor state, becomes when $\tau_{j1,\ldots,jn}$ is applied in a state $u$. When $c-\tau_{j1,\ldots,jn}(u)$ holds, this is expressed generally by a conditional equation denoted by the keyword *ceq* . The value returned by $o_{i1,\ldots,im}$ is not changed if $\tau_{j1,\ldots,jn}$ is applied in a state $u$ such that $not(c-\tau_{j1,\ldots,jn}(u))$.

In table 1, we can see Social OTS in CafeOBJ notation. $H$ and $H'$ are CafeOBJ variables for the hidden sort of the base-line objects and the composite object respectively. $V_{ij}$ denote CafeOBJ variables for the visible sorts $D_{ij}$ and finally $V_1$ is a variable for the visible sort that represents the unique identification of base-line object, say ID.

**Table 1.** Social Network OTS in CafeOBJ terms.

| OTS Observers | CafeOBJ Notation |
|---|---|
| regular observers; $o : Y' \to D$ | `bop o : V`$_{i1}$` ... V`$_{im}$` H' -> V.` |
| projection observers; $p : Y' \to Y$ | `bop o : V`$_{i1}$` ... V`$_{im}$` H' -> H.` |
| regular transitions; $\tau : Y' \to Y'$ | `bop τ : V`$_{i1}$` ... V`$_{im}$` H' -> H'.` |
| special transitions; $a : Y' \to Y'$, $d : Y' \to Y'$ | `bop a : V`$_1$` H' -> H'.`<br>`bop d : V`$_1$` H' -> H'.` |
| effective conditions; $c_{\tau j1,\ldots,jn}$ | `op c-τ`$_{j1,\ldots,jn}$` : V`$_{i1}$` ... V`$_{im}$` H' -> Bool.` |

## 4  An algebraic specification of a Social Network

As described in section 1, there are many instantiations of the abstract concept of a Social Network, but all of them comply with the abstract definition of the Social Network OTS we gave. When it comes to creating a concrete specification, we decided that it is more important to present the one corresponding to Social Network Sites due to their increasing popularity, impact on everyday life and rich application extent.

The building block of a social network site system is the concept of a *User Profile*. User Profiles, in a sense define the user behind them via a collection of data types like; *photo albums*, a *wall space*, their *inbox* etc. Finally, User Profiles are connected to each other with a (at times associative) relationship, usually called *'friendship'*.

In the OTS/CafeOBJ approach these data types need to be defined in separate modules and then imported to the module defining the User Profile. The abstract data type of the wall, represents an open space for the users (including the profile owner) to publish content on the Profile web page of the user. Content is a triplet; the identity of the user that published it, a unique identifier for the content and finally the content itself. The CafeOBJ module defining the wall space can be seen below. The rest of the data types are defined in a similar way.

```
mod! WALLCONTENT { pr(CONTENT + ACCOUNTID+Nat + LIST3)
[Wallcontent]
op _&_&_ : Accountid Nat Content -> Wallcontent
op Id? : Wallcontent -> Nat
op Accountid? : Wallcontent -> Accountid
op Like? : Wallcontent -> ListofAccountid }
```

### 4.1 Profile OTS

The *Profile* is the CafeOBJ module that specifies the user profiles. This module defines an OTS, whose state space is declared by the hidden sort *ProfileSys*. As described in previous sections an OTS is a collection of Observers and Transitions. In tables 2 and 3 we present the informal definition, OTS notation and OTS/CafeOBJ definitions of some of the observers and transitions of the Profile OTS.

**Table 2.** Observers defining the Profile OTS.

| Informal description | OTS observers | CafeOBJ code |
|---|---|---|
| The wall space | $wall : Y \to D_1$ | `bop wall : ProfileSys -> Walllist` |
| The inbox mail service | $inbox : Y \to D_1$ | `bop inbox : ProfileSys -> Inboxlist` |
| A collection of photos | $photoalbum : Y \to D_1$ | `bop photoalbum : ProfileSys ->Photolist` |
| A set of user ids that have endorsed this post/photo/etc | $likeset : Y \ D_1 \ D_2 \ D_3 \to D_4$ | `bop likeset : ProfileSys Nat Placeholder -> Setofaccountid` |
| A list of user ids that are connected with the profile through the friendship relation | $friends : Y \to D_1$ | `bop friends : ProfileSys -> ListofAccountid` |
| Denotes whether a profile is private or not | $visibility : Y \to D_1$ | `bop visibility : ProfileSys -> Bool` |
| A identification for the profile which is unique | $myid : Y \to D_1$ | `bop myid : ProfileSys -> Accountid` |

**Table 3.** Transitions defining the Profile OTS.

| Informal description | OTS transitions | CafeOBJ code |
|---|---|---|
| Receive a friend request from a user | $receivefriend : Y\ D_1 \to Y$ | `bop receivefriendrequest : ProfileSys Accountid -> ProfileSys` |
| Accept a friend request from a user | $acceptfriend : Y\ D_1 \to Y$ | `bop acceptfriendrequest : ProfileSys Accountid -> ProfileSys` |
| A user endorses a post/photo/etc | $reclike : Y\ D_1\ D_2\ D_3 \to Y$ | `bop receivelike : ProfileSys Placeholder Nat Accountid -> ProfileSys` |
| A user checks your list of friends | $viewfriends : Y\ D_1 \to Y$ | `bop viewfriends : ProfileSys Accountid -> ProfileSys` |

Apart from defining a transition rule, we need to define what each observer observes in the initial state of the system and what when each transition is applied in an arbitrary state. This is done by first defining in CafeOBJ terms the effective condition for each rule. Next we write (in equation form) the observed value, of each observer for the new system state when that effective condition holds. Finally, we need to sate explicitly in an equation that when the effective condition does not hold the state of the system remains the same.

### 4.2 Social Network OTS

Having described the *Profile* entity as an OTS in CafeOBJ, we now create the specification of a Social Network. Our approach of modeling a social network is that of a dynamic behavioral object as we described in sections 2 and 3. It is clear that a social network should be modeled as a system that dynamically creates/deletes other systems (the profiles). The composite object will represent the Social Network that contains as base-line objects dynamically many Profile OTS systems. The definition of the social network OTS in terms of observers and transitions can be seen below:

**Table 4.** Social Network OTS transitions

| Informal description | OTS transitions | CafeOBJ code |
|---|---|---|
| Install a new Profile with a unique id | $add : Y\ D_1 \to Y$ | `bop add : Sys Accountid -> Sys` |
| Delete a specific profile | $delete : Y\ D_1 \to Y$ | `bop del : Sys Accountid -> Sys` |
| Profile with say id1, receives some content from a profile, say id2. | $receiveSN : Y\ D_1\ D_2\ D_3\ D_4 \to Y$ | `bop receiveSN : Sys Accountid Content Accountid Placeholder -> Sys` |
| Profile, id1, accepts a friend request from | $acceptfriendSN : Y\ D_1\ D_2 \to Y$ | `bop acceptfriendSN: Sys Accountid` |
| profile id2 | | Accountid -> Sys |
| The user behind profile id2, sees the photos of profile id1 | $viewphotoSN : Y\ D_1\ D_2 \to Y$ | bop viewphotoSN : Sys Accountid Accountid -> Sys |

**Table 5.** Social Network OTS observers.

| Informal description | OTS observers | CafeOBJ code |
|---|---|---|
| Set of installed profiles of the social network | $accounts : Y \to D_1$ | op accounts : Sys -> Setofaccountid |
| Projection from the state of the network to that of a profile | $profile : D_1\ Y \to Y'$ | op profile : Accountid Sys -> ProfileSys |

Transition rules add and del hold special interest, as they are responsible for creating and deleting profiles. The effects of transition *add* are defined by the following equations:

```
1. eq c-add(A1,S)  = not ( A1 /in accounts(S) ) .
2. ceq profile(A2,add(A1,S))= init if (A1 = A2) and c-add(A1,S) .
3. ceq profile(A2,add(A1,S))= profile(A2,S)  if (not (A1 = A2)) and c-add(A1,S).
4. ceq accounts(add(A1,S))= (A1 U accounts(S)) if c-add(A1,S).
```

Line 1 is the effective condition of the *add* transition rule and states that in order to add a profile in a social network, this profile must not belong to the network already. In lines 2 and 3 we describe how the state of the profile A2 is affected when we are adding the profile A1. More precisely, line 2 says that if A1 is the same with A2 and the effective condition holds then the state of A2 will be the initial sate, as that is defined in the profile OTS. This was a *Projection*, because from a state of the composite OTS namely *add(A1,S)* we derived a state of the component object, namely *init*. In line 3 we say that if the two profiles are different the state of A2 is not affected, as it is expected. Finally, in line 4, we define that the set of all installed profiles will contain A1 as well if the effective condition holds. The rest of the transition rules correspond directly through projections to transitions rules of the component OTS objects.

It is also interesting to see how the change in the state of one component object can change the state of another component object.

Assume that profile with account id A1 accepts a friend request from account id A2. It is easy to note that the above intuitively corresponds to a change of state for the component OTS of profile A1. But the friendship relationship is *reflexive* in our specification, meaning that if user A2 is added to the friends of A1, then automatically A1 should be added to the friends of A2. So, *a change in the state of one component OTS can change the state of another.* Consecutively, the observers for the sates of both Profile OTSs need to change, but only them.

```
1. ceq profile(A3 ,acceptfriendSN(A1,A2,S))= accept
friendrequest(A2,profile(A1,S)) if c-acceptfriendSN
(A1,A2,S) and (A1 = A3) .
2. ceq profile(A3,acceptfriendSN(A1,A2,S)) = accept
friendrequest(A1,profile(A2,S)) if c-acceptfriendSN
(A1,A2,S) and (A1 = A3) .
3. ceq profile(A3,acceptfriendSN(A1,A2,S))= profile
(A3,S) if c-acceptfriendSN(A1,A2,S) and not (A1 = A3) .
```

## 5   Verification

One of the main advantages of our approach is the ability to verify that the system preserves several critical *safety* or *liveness* properties. This is a feature of most specification languages. In this paper we have chosen to use CafeOBJ and to prove our properties in a semi-automated way, using the Proof Score method [12] which is described later in this section. . A property for a system is said to be *invariant* if it holds in any given reachable state of the system. In order to demonstrate the proving power of our approach we have chosen to show that the following two invariant properties hold in our specification

| Invariant 1 | *It is not possible for someone that is not your friend to see your photos, if your profile is private.* |
|---|---|
| Invariant 2 | *If your profile is private, a user that is not your friend can not see the list of your befriended users.* |

In the rest of this section we will display the proving procedure for the first invariant property. In order to prove an invariant property using the CafeOBJ/OTS method, we have to follow several steps [12].

We express the property we want to prove formally as a predicate , say *invariant pred(p,x)*, where $p$ is a free variable for states and $x$ symbolizes other free variables of *pred*. Then, we write *pred(p,x)* using CafeOBJ terms in a module, say INV.

```
op inv : Sys Accountid Accountid Picture -> Bool
eq inv(S,A1,A2,Pi) = (not(visibility(profile(A1,S)))
or not(A2 //in friends(profile(A1,S))))
implies not(view(profile(A1,S),A2,Pi)) .
```

We express the predicate in each inductive case, using two constants $s, s'$ that denote any state and the successor state after applying a transition rule to that state, in the module ISTEP. i.e the induction step.

```
op istep : -> Bool
eq istep = inv(s,a1,a2,pi) implies inv(s',a1,a2,pi).
```

We prove that the predicate holds for any initial state, say *init*, by reducing *pred(init,x)*.

```
open INV
red inv(initSN,a1,a2,pi) .
close
```

Next we write a proof score for each inductive case. If it is reduced to true, it is proven that the transition rule preserves *pred(p, x)* in this case. Otherwise, we may have to split the case, or need to discover lemmas, or we may prove that the predicate *is not invariant* to our system. For the case of the viewphotoSN transition, we have the following:

```
open ISTEP
eq s' = viewphotoSN(a1,a2,s) .
red istep .
close
```

In this case CafeOBJ returns neither true nor false. Thus, we have to split the case to help the system with the reduction. The most natural decision is to split the effective condition of the transition based on whether it holds or not.

```
open ISTEP
eq c-viewphotoSN(a1,a2,s) = false .
eq s' = viewphotoSN(a1,a2,s) .
red istep .
close
```

The above proof passage refers to the case that the effective condition of the transition does not hold and CafeOBJ returns true. Now we have to check the case where the effective condition holds. Here, we have replaced the equation eq c-receivefriendSN (a1,a2,s) = true with its definition, that is the two equations in italic.

```
open ISTEP
-- eq c-receivefriendSN(a1,a2,s) = true .
```
*eq (a1 /in accounts(s)) = true .*
*eq (a2 /in accounts(s)) = true .*
```
eq s' = receivefriendSN(a1,a2,s) .
red istep .
close
```

CafeOBJ returns again neither true nor false. Thus, we decided to split the effective condition of the transition rule of the component OTS. In the case where the effective condition is false CafeOBJ returned true. As previously, we must continue with the case where the effective condition is true. The two equations in italic correspond again to the definition of the effective condition for this transition.

```
open ISTEP
-- eq c-viewphotoSN(a1,a2,s) = true .
eq (a1 /in accounts(s)) = true .
```

```
eq (a2 /in accounts(s)) = true .
eq (a2 //in friends(profile(a1,s)) ) = false .
-- eq c-viewphoto(profile(a1,s) ,a2) = true .
eq visibility(profile(a1,s)) = true .
eq (a2 //in friends(profile(a1,s))) = true .
eq s' = viewphotoSN(a1,a2,s) .
red istep .
close
```

CafeOBJ returns true for the above proof passage and hence, this proves our safety property for the case of the *viewphotoSN* transition rule. Following the same procedure, CafeOBJ returned true for all transitions and thus our proof concludes. The second invariant property was also formally proven by using again the CafeOBJ/OTS method.

## 6  Future applications

We have presented the definition of a Social Networking Service as time evolving system that is consisted of other evolving systems that interact with each other, change each other states and as a consequence the state of the whole Social Networking Service. This was formally defined as a dynamic behavioral object through the OTS approach (and defined the Social Network OTS). With the help of CafeOBJ, an algebraic specification language, we have provided a specification for an implementation of a Social Networking Site that complies with the definition of the Social Network OTS. Finally, we have demonstrated how various safety properties can be verified using the OTS/proof score methodology.

This is a first approach to formally describe Social Networks using formal methods and in particular algebraic specification techniques. We believe that we have argued adequately for the advantages of this approach.

This work can be expanded in order to fully specify a real-life online Social Network implementation. For example, a profile can be either public or private in our specification, whereas there are more levels of privacy in-between. A user can choose what parts of his profile wants public or can select specific users that can see all of the profile, or exclude specific users from some photos. In addition, handling more forms of content can be added to this specification, such as video, game applications etc. Also, many other critical properties can be verified. For example that a third party program cannot access your personal data unauthorized.

Finally, we believe that a full specification using this approach ideally should be used in the pre-coding stage of development as means to verify that the design holds the desired critical properties.